\documentclass[twocolumn]{aastex631}
\usepackage{graphicx}   
\usepackage{lipsum} 
\usepackage{amsmath}

\usepackage{color}

\def\hmpc{h^{-1}{\rm Mpc}}

\def\hkpc{h^{-1}\, {\rm kpc}}

\def\hmsun{{h^{-1} M_{\odot}}}
\def\msun{\, M_{\odot}}
\def\mbh{\, M_{\rm BH}}
\def\mstar{\, M_{*}}

\def\astrid{\texttt{ASTRID} }

\shorttitle{Merger of triple quasars}
\shortauthors{Y.Ni et al.}
\graphicspath{{./}{figures/}}

\begin{document}

\title{
Ultramassive black holes formed
by triple quasar mergers at $z\sim 2$}

\author{Yueying Ni}
\affiliation{McWilliams Center for Cosmology, Department of Physics, Carnegie Mellon University, Pittsburgh, PA 15213}
\affiliation{Center for Astrophysics $\vert$ Harvard \& Smithsonian, Cambridge, MA 02138, US}

\author{Tiziana Di Matteo}
\affiliation{McWilliams Center for Cosmology, Department of Physics, Carnegie Mellon University, Pittsburgh, PA 15213}

\author{Nianyi Chen}
\affiliation{McWilliams Center for Cosmology, Department of Physics, Carnegie Mellon University, Pittsburgh, PA 15213}

\author{Rupert Croft}
\affiliation{McWilliams Center for Cosmology, Department of Physics, Carnegie Mellon University, Pittsburgh, PA 15213}

\author{Simeon Bird}
\affiliation{Department of Physics \& Astronomy, University of California, Riverside, 900 University Ave., Riverside, CA 92521, US}

\correspondingauthor{Yueying Ni}
\email{yueyingn@andrew.cmu.edu}

\begin{abstract}
The origin of rare and elusive ultramassive black holes (UMBH, with $\mbh > 10^{10} \msun$) is an open question.
Using the large volume cosmological hydrodynamic simulation \astrid, we report on the formation of an extremely massive UMBH with $\mbh \sim 10^{11} \msun$ at $z \sim 2$. 
The UMBH is assembled as a result of two successive mergers of massive galaxies each with stellar mass $\mstar > 3 \times 10^{11}$ that also produces a bright, rare triple quasar system powered by three $\sim 10^9 \msun$ black holes.
The second merger of supermassive black holes (SMBHs) follows the first after 150 Myrs. 
The merger events lead to sustained Eddington accretion onto the central SMBH, forming an UMBH in the center of a massive compact stellar core with $\mstar > 2 \times 10^{12} \msun$. 
The strong feedback of the UMBH quenches the surrounding star formation to $< 10 \msun$/yr in the inner 50 $\hkpc$ region. 
There are two more UMBHs with $\mbh > 5 \times 10^{10} \msun$ at $z>2$ in \astrid which are also produced by major mergers of galaxies, and their progenitors can be observed as quasar triplets of lower luminosity.
The rarely observed quasar multiples can be the cradle of UMBHs at high redshift, and likely end up in the center of the most massive clusters.

\end{abstract}


\section{Introduction}
\label{section1:introduction}








Probing the most massive end of the SMBHs and their relation with host galaxy properties is crucial for us to reach a comprehensive understanding of how they grow and co-evolve with cosmic structures like galaxies, galaxy groups and even clusters of galaxies. 
Over the last decade, observations of the local universe have established the existence of a few UMBHs (with $\mbh > 10^{10} \msun$) in some bright cluster galaxies (BCGs) \citep[e.g., see][]{McConnell2011Natur.480..215M, Hlavacek-Larrondo2012MNRAS.424..224H}.
The current most massive black hole with direct dynamical mass measurement is $\mbh \sim 4 \times 10^{10} \msun$ \citep{Mehrgan2019ApJ...887..195M} at the centre of Holm15A, the central galaxy of Abell 85.
Indirect mass measurements of high redshift quasars suggest the existence of UMBHs with $\mbh > 6 \times 10^{10} \msun$ \citep[e.g., TON618,][]{Shemmer2004ApJ...614..547S}.
Given the difficulties and uncertainties lying in the mass measurement of SMBHs \citep[e.g., see][for a review]{Kelly2012AdAst2012E...7K,Peterson2014SSRv..183..253P}, it still remains unclear whether there exist or can exist UMBHs with larger mass.
Some theoretical studies suggest that there is an upper limit for black hole mass in the  $M_{\rm BH,max} = 5 \times 10^{10} \sim 2 \times 10^{11} \msun$ regime, above which they cannot grow through luminous accretion of gas \citep{Natarajan2009MNRAS.393..838N,King2016MNRAS.456L.109K}.

It has been suggested that UMBHs could be remnants of extremely luminous quasars seen at higher redshift and hence may form around the peak of the quasar phase at $z\sim 2$.
Major mergers of gas-rich galaxies at this epoch fuel quasars and give rise to the most extreme black hole growth and assembly.

The existence of multiple, simultaneously active, SMBHs, in galaxy mergers represents a key observational test 
\citep[e.g.][]{Bennert2008ApJ...677..846B} for understanding the processes regulating quasar activity and the growth of SMBHs.
Extreme examples of these are the rare quasar multiples, that are rather challenging to detect due to their rarity, the required angular resolution, incompleteness, interlopers of lensed pairs, etc.
There are, however, two observations of (luminous) quasar triplets that have been reported so far, QQQJ1432-0106 at $z=2.1$ observed by~\cite{Djorgovski2007ApJ...662L...1D}, and QQQ J1519+0627 at $z=1.5$ reported by~\cite{Farina2013MNRAS.431.1019F}.
The two quasar triplets are observed on galactic scales with separations around tens to hundreds of pkpc at $z \sim 2$.  
Based on the velocity differences in the triplet, those authors propose that the three quasars are `caught in the act' and forming a physical structure of mass $10^{13} \msun$.
Based on the rarity of this event and the mass of the remnant galaxy and black hole, such a triple quasar system may thus be an ideal candidate for a progenitor of an UMBH in a central cluster galaxy today.

The rare quasar multiples and their mergers are difficult to model in cosmological simulations due to the limited volume.
In this work, we inspect the formation of rare UMBHs predicted by the \astrid simulation at $z \geq 2$ and investigate their origin.
\astrid is a recently developed large volume, high-resolution cosmological hydrodynamic simulation \citep{Ni-Asterix,Bird-Asterix}.
Its large volume of (250 $\hmpc$)$^3$ (the greatest volume for a galaxy formation simulation to date) allows a systematic study of the rare quasar and galaxy population at cosmic noon, and can probe the most extreme events such as multiple major mergers of massive galaxies and systems of quasar multiples.

The paper is organized as follows. 
We briefly introduce the \astrid simulation in Section~\ref{section2:Method}.
In Section~\ref{section3:Result}, we give \astrid predictions for the statistics of the dual and triple quasars with galactic separations. 
We describe in detail how the merger of the brightest triple quasar system forms an UMBH with mass $\mbh > 10^{11} \msun$ and explore the various host galaxy properties.
We discuss the result in Section~\ref{section4:Discussion} and conclude in Section~\ref{section5:Conclusion}.

\section{Simulation}
\label{section2:Method}

The \astrid simulation is a cosmological hydrodynamical simulation performed using a new version of the Smoothed Particle Hydrodynamics code \texttt{MP-Gadget}.
The simulation evolved a cube of $250 \hmpc$ per side with $2\times5500^3$ initial tracer particles of dark matter and baryons, and has currently reached $z=2$.
It is the largest cosmological simulation of galaxy formation to date that covers the epoch of cosmic noon.

\astrid achieves a dark matter particle mass resolution of $M_{\rm DM} = 6.7 \times 10^6 \hmsun$ and $M_{\rm gas} = 1.3 \times 10^6 \hmsun$ in the initial conditions.
The gravitational softening length is $\epsilon_{\rm g} = 1.5 \hkpc$ for both DM and gas particles.
The simulation implements a variety of sub-grid models for physics governing the formation of galaxies and SMBHs and their associated supernova and AGN (Active Galactic Nuclei) feedback, inhomogeneous hydrogen and helium reionization, and the effect of massive neutrinos.
We refer the readers to the introductory papers \cite{Bird-Asterix,Ni-Asterix} for detailed descriptions of physical models used in the simulation.

Galaxies in the simulation are identified through \texttt{SUBFIND} \citep{2001MNRAS.328..726S} in a post-processed manner. At $z=2$, \astrid contains a statistical sample of very massive galaxies and bright quasars, with about $3 \times 10^{3}$ galaxies having $\mstar>10^{11} \msun$ and 13 galaxies having $\mstar \geq 10^{12} \msun$. 
It produces 709 black holes with $\mbh > 10^{9} \msun$, three of which have $\mbh > 5 \times 10^{10} \msun$, with the most massive one having reached $10^{11} \msun$. 
The large volume of \astrid provides us with an ideal suite to study the galaxy and quasar population at cosmic noon.


\section{Results}
\label{section3:Result}
\subsection{Quasar multiples at the bright end of the quasar population}

\begin{figure}
\centering
  \includegraphics[width=1.0\columnwidth]{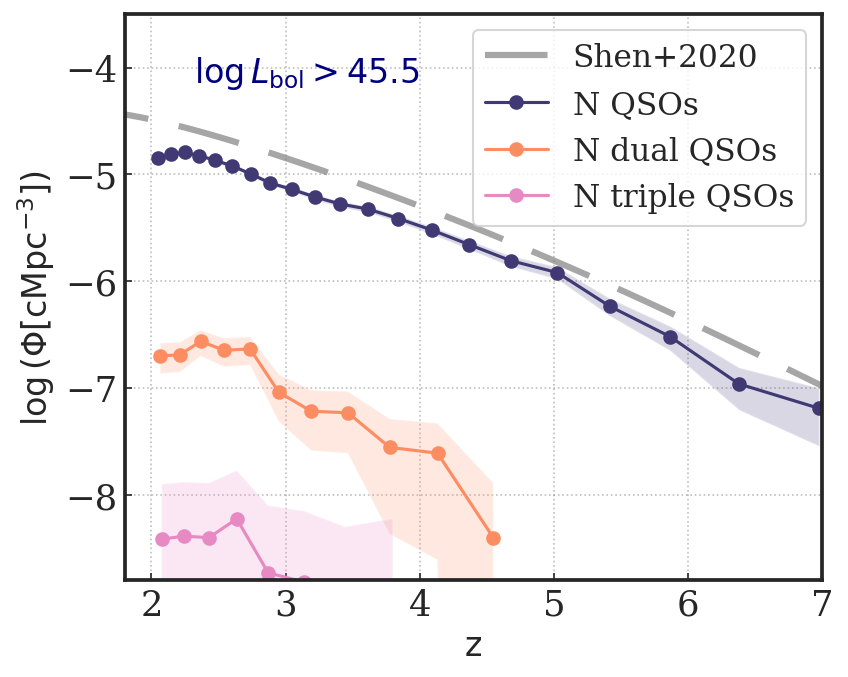}
  \caption{
  Cumulative number density of quasars as well as dual and triple quasars as a function of redshift. 
  The gray dashed line shows the observational determination of the quasar number density from \citet{Shen2020MNRAS.495.3252S}, obtained by integrating the QLF fit from a collection of all-band quasar observations. 
  The dark blue line shows the number density of quasars with $\log L_{\rm{bol}} > 45.5$ within the 250 $\hmpc$ volume of \astrid. 
  The orange (pink) line shows the (time-averaged) number density of quasar pairs (triples) where quasars are within $5 \hkpc < r < 200 \hkpc$ from each other. 
  The shaded area corresponds to the Poisson noise. 
  }
  \label{fig:Nqso-z}
\end{figure}


We start with a brief overview of the quasar population predicted by \astrid, focusing on the bright quasars that can be probed by observations of dual and triple quasar systems. 
Figure~\ref{fig:Nqso-z} shows the quasar population with luminosity threshold $L_{\rm bol} > 10^{45.5}$ ergs/s \citep[commensurate with observations of e.g.,][]{Shen2022arXiv220804979S}: \astrid has a sample of $700 \sim 800$ quasars, in broad agreement with the observational estimate of \cite{Shen2020MNRAS.495.3252S} over the full redshift range. 
Given the transitory nature of the active quasar phase, we calculate the number density of dual and triple quasars in a time-averaged manner (as shown by the orange and pink lines).
Only $\sim 1\%$ and $0.02\%$ of quasars are in dual and triple quasar systems respectively, with separation from $5 < r < 200 \hkpc$ (corresponding to angular separation $\delta_{\theta} = 0.3 \sim 10$\arcsec at $z \sim 2$).
Thereby, based on time averaged estimation, \astrid predicts dual quasar fraction $f_{\rm QQ} \sim 1 \times 10^{-2}$ and triple quasar fraction $f_{\rm QQQ} \sim 2 \times 10^{-4}$ amongst quasars with $L_{\rm bol} > 10^{45.5}$ ergs/s at $z = 2 \sim 3$.

With the large volume of \astrid, we are able to find a handful of rare quasar triplets at $z\sim 2$.
In the redshift range of $z>2$, there exist 5 such systems in \astrid that contain three quasars that meet the luminosity and distance criteria at a given time, and two of those triplet systems only last for a short period of time with $t_{\rm QQQ} < 50$ Myr.
In the entire simulation, there is only one triplet in which all three quasars have $L_{\rm bol} > 10^{46}$ ergs/s. 
Remarkably, we find that this system is the progenitor of the $10^{11} \msun$ black hole, the most extreme UMBH formed in the simulation.
We focus on this brightest quasar triplet and show how the two subsequent mergers of the quasar host galaxies allow the mass assembly of this extreme UMBH.

\subsection{Massive mergers and triple quasars}

\begin{table*}
\centering
\begin{tabular}{llllll}
    \hline
    & redshift & $M_{\rm BH}$ [$\msun$] & $L_{\rm bol}$ [ergs/s] & $M_*$ [$\msun$] & $M_{\rm halo}$ [$\msun$] \\
    \hline
    & $z=2.7$ & [$3.0\rm{e}9$, $8.3\rm{e}8$, $5.2\rm{e}8$] & [$2.6\rm{e}{46}$, $1.7\rm{e}{46}$, $9.6\rm{e}{45}$] & [$4.6\rm{e}{11}$, $5.5\rm{e}{11}$, $3.2\rm{e}{11}$] & $6.3\rm{e}{13}$ \\
    
    before merger 1 & $z=2.5$ & [$4.3\rm{e}9$, $2.3\rm{e}9$, $1.6\rm{e}9$] & [$1.0\rm{e}{47}$, $1.6\rm{e}{46}$, $5.0\rm{e}{46}$] & [$9.8\rm{e}{11}$, $4.7\rm{e}{11}$, $3.6\rm{e}{11}$] &  $7.8\rm{e}{13}$ \\
    
    after merger 1 & $z=2.4$ & [$9.4\rm{e}9$, $2.7\rm{e}9$] & 
    [$1.2\rm{e}{47}$, $1.8\rm{e}{46}$] & 
    [$1.6\rm{e}{12}$, $5.0\rm{e}{11}$] & $9.0\rm{e}{13}$ \\
    
    after merger 2 & $z=2.3$ & [$7.2\rm{e}{10}$] & 
    [$4.4\rm{e}{48}$] & 
    [$2.5\rm{e}{12}$] & $9.4\rm{e}{13}$ \\
    
    & $z=2.0$ & [$1.8\rm{e}{11}$] & 
    [$1.9\rm{e}{45}$] & 
    [$2.3\rm{e}{12}$] & $1.1\rm{e}{14}$ \\
    
    \hline
\end{tabular}
\caption{Host information for the triple quasar system from $z=2.7$ to $z=2.0$. $M_{\rm halo}$ is the halo mass of the FOF group.}
\label{tab:table1}
\end{table*}

\begin{figure*}
\centering
  \includegraphics[width=1.0\textwidth]{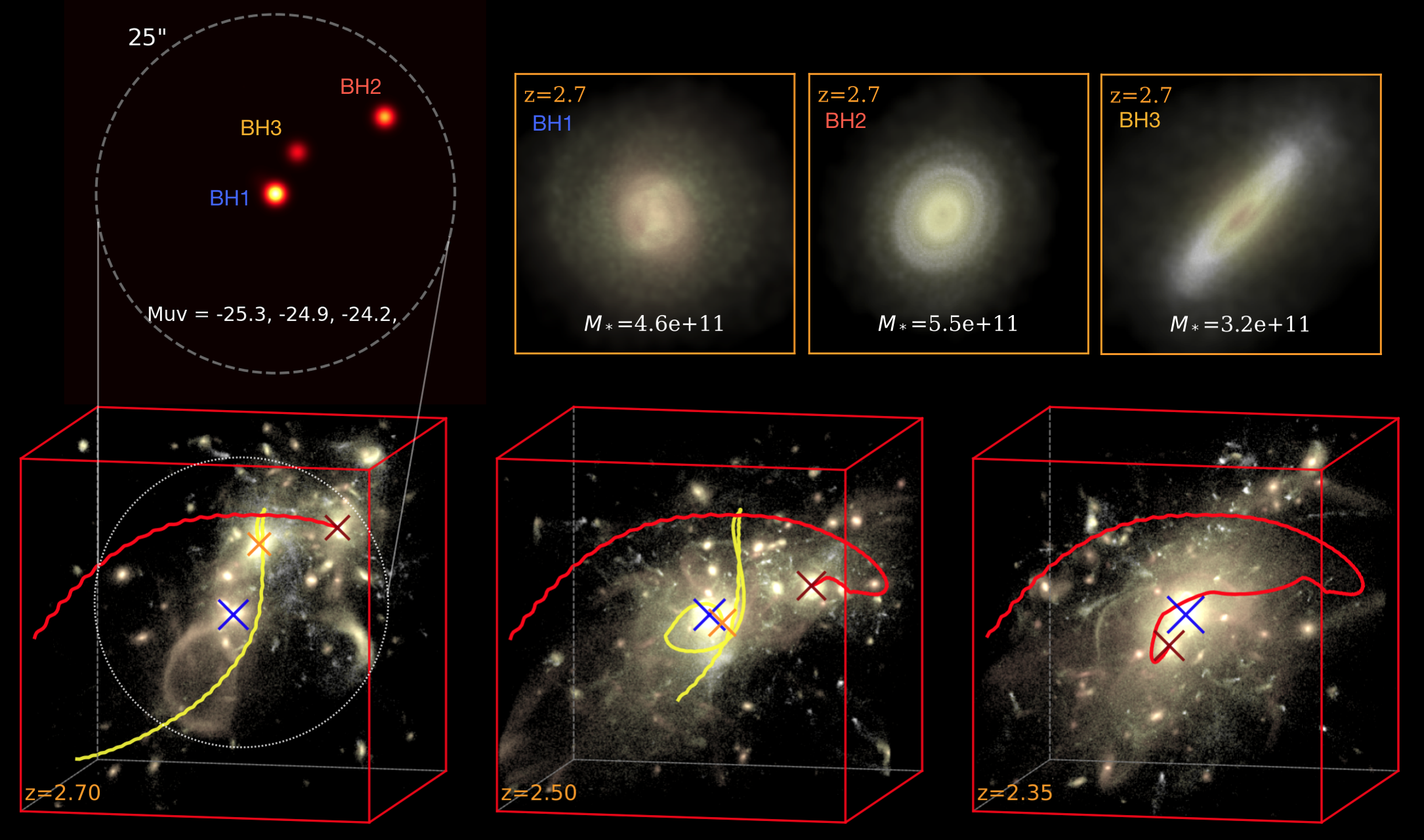}
  \caption{Illustration of the quasar triplet system and its environment (host galaxies). 
  The background in the lower panels shows the stellar density field in a 600 $\hkpc$ region centered at the most massive quasar (BH1). 
  Crosses mark the positions of the three quasars. 
  The red and yellow lines mark the trajectories of the other two quasars (BH2 and BH3) in the reference frame of BH1. 
  The left upper panel shows a cartoon of the quasar triplet system in a field of view with 25$\arcsec$ (500 $\hkpc$ at $z=2.7$).
  The image is convolved with a Gaussian point-spread function with full-width at half-maximum $0.5 \arcsec$.
  The right upper panel illustrates the host galaxies for the three quasars at $z=2.7$ in 20 $\hkpc$ regions, projected in the same direction as the cartoon on the left. Colors show the stellar age, with older stars being redder. 
  }
  \label{fig:illustration}
\end{figure*}

\begin{figure*}
\centering
  \includegraphics[width=\textwidth]{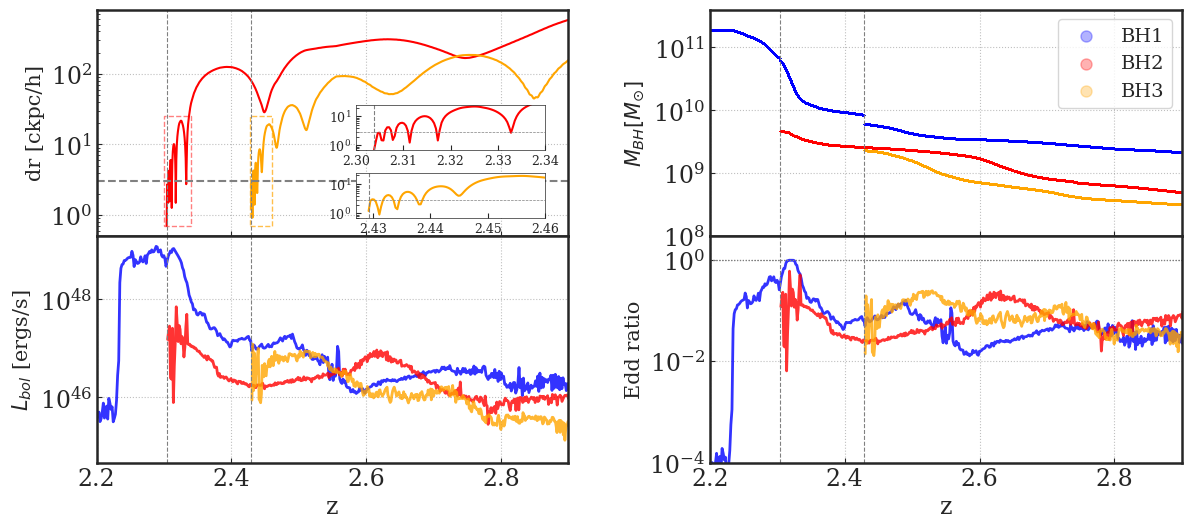}
  \caption{ 
  Detailed evolution of each of the three quasars. The blue line represents the BH1, the central black hole in Figure~\ref{fig:illustration}. The red (orange) line in the first panel gives the 3D distance between BH2 (BH3) and BH1. The horizontal dashed line marks $2 \epsilon_g$ as the gravitational softening when the black hole merger occurs in the simulation. The two insets zoom into the dashed rectangles to better show the evolution before the SMBH mergers. 
  The bottom left, top right, and bottom right panels show the evolution of bolometric luminosity, black hole mass and the Eddington ratio respectively. 
  The vertical dashed lines in each panel mark the time of SMBH mergers. 
  }
  \label{fig:bh-detail}
\end{figure*}

\begin{figure*}
\centering
  \includegraphics[width=\textwidth]{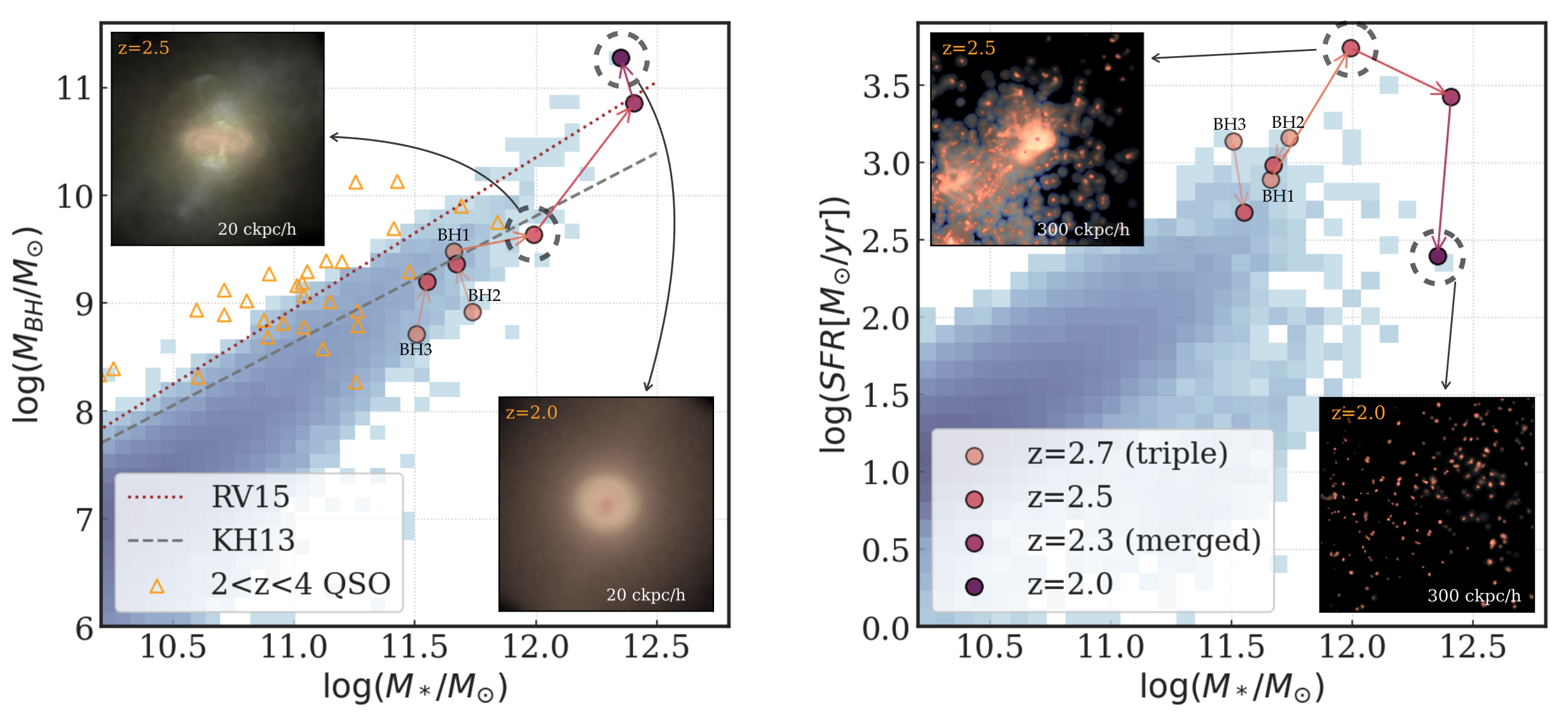}
  \caption{
  Relation between the stellar mass of galaxies with (i) mass of the most massive black hole in that galaxy (left panel) and (ii) star formation rate of the galaxy (right panel). 
  The blue background shows the 2D histogram for all the galaxy populations at $z=2$ in the \astrid simulation. 
  The gray dashed line and brown dotted line give the observational fitting of $\mbh - \mstar$ from \citet{Kormendy2013,Reines2016} based on $z\sim0$ AGN observations.
  The orange data points give $2<z<4$ quasars collected from \citet{Kormendy2013}.
  The circles mark the evolution of the triple quasar system from $z=2.7$ to $z=2.0$.
  Insets in the left panel illustrate the morphology of the BH1 host galaxy (in face-on direction) at $z=2.5$ and $z=2.0$ in the $20 \hkpc$ region. 
  Insets in the right panel illustrate the gas density field colored by the star formation rate (with red color indicating high star formation rate) in the $300 \hkpc$ region centered at BH1. 
  }
  \label{fig:Scaling-Relation}
\end{figure*}

The triple quasar system is formed at the early stages of a merger of three massive galaxies, each hosting strongly accreting SMBHs.  
The first merger (merger 1) is closely followed by a second one (merger 2).

Table~\ref{tab:table1} summarizes the properties of the galaxies and black holes taking part in this event.
The system is hosted by one of the largest halos in the simulation, with mass $M_{\rm halo} \sim 6-9\times 10^{13} \msun$.
In the initial phase of the interaction, the three host galaxies all have a stellar mass of a few $10^{11} \msun$.
The system starts from a stage of quasar triplet at $z \sim 2.7$, with each quasar having $L_{\rm bol} \sim 10^{46}$ erg/s, and the associated 3 SMBHs (with masses ranging from $0.5-3\times 10^{9} \msun$) residing in their respective host galaxies.
This stage is followed by two subsequent mergers of galaxies and the SMBHs, leaving the final remnant massive galaxy hosting an UMBH of mass $M_{\rm BH} \sim 10^{11} \msun$ after the second merger at $z \sim 2.3$.

To illustrate those stages, Figure~\ref{fig:illustration} shows the triple quasars together with their respective host galaxies and snapshots over the dynamical evolution of this system. 
The evolution of the black hole masses and luminosities at each stage of the evolution is given in Figure~\ref{fig:bh-detail}.
The triple quasars at $z=2.7$ is shown at angular separations of 9\arcsec and 3\arcsec for BH1-BH2 and BH1-BH3 respectively. 
These are comparable to those in the two known quasar triplets \citep{Farina2013MNRAS.431.1019F}.
The quasar triplet is thus associated with the three SMBHs, each residing in massive host galaxies (see Table~\ref{tab:table1}) that are in the initial phase of interacting but have not yet merged with each other. 
The three quasars all have $L_{\rm bol} > 10^{46}$ ergs/s, being the brightest quasars (triplet) in the entire simulation.

The phase of the quasar triplet is then followed by two major mergers of the SMBHs as well as their host galaxies. 
The two black hole merger events occur within 150 Myr; the first merger of BH2 and BH1 takes place at $z=2.43$ and the second merger of BH3 and BH1 takes place at $z=2.304$. 
The upper left panel of Figure~\ref{fig:bh-detail} shows the distance of BH2 and BH3 from the central BH1 as they inspiral and eventually merge (which occurs within $2\epsilon_g$).
We define the merger elapse time $t_{\rm elapse}$ as the time interval between the first time when two black holes are within $2\epsilon_g$ and the time when they numerically merge.
$t_{\rm elapse}$ for the first and second merger are 11 Myrs and 36 Myrs respectively. 
The time taken between pairing and merger for these two black holes is short compared to the overall $t_{\rm elapse}$ distribution for all black hole merger events in \astrid, which peaks at $t_{\rm elapse} \sim 200$ Myr \citep[see][]{Ni-Asterix}.
The short merger elapse time is a result of the high stellar density surrounding the black holes and associated effective dynamical friction in these environments which quickly dissipates the kinetic energy of the black holes.

At around $z = 2.3$,  just after the final black hole merger, we have the most luminous quasar in the entire simulation, with $L_{\rm bol} > 10^{48}$ ergs/s (Figure~\ref{fig:bh-detail}).
This critical accretion phase with sustained Eddington rate grows the black hole mass by about tenfolds, and is induced by the major merger of the host galaxy. 
During this most active accretion phase, the central black hole is however surrounded by high density gas with line-of-sight column density ranging from $N_{\rm H} = 5 \times 10^{23} \sim  10^{25}$ cm$^{-2}$ that heavily obscuring most of the sightlines to the quasar.
The close-to-Eddington phase lasts for about 140 Myrs and is finally quenched by powerful AGN feedback. 
At $z<2.23$ (96 Myr after the second merger event), powerful gas outflow driven by the AGN feedback clears out the dense gas in its surroundings and quenches the active accretion. 
The luminosity of the UMBH decreases to $10^{46}$ ergs/s and the corresponding Eddington ratio falls into the $\lambda_{\rm Edd} \sim 10^{-4}$ regime.

\subsection{The remnant host galaxy of a triple quasar and newly formed UMBH }


Figure~\ref{fig:Scaling-Relation} shows the $\mbh$ - $\mstar$ and SFR - $\mstar$ relations for all the $\mstar > 10^{9} \msun$ galaxies in \astrid at $z=2$ and highlights the triple merger system and its evolution from $z=2.7$ to $z=2.0$, in these two planes.
Insets to Figure~\ref{fig:Scaling-Relation} illustrate the evolution of the BH1 host galaxy and the star formation environment in its surroundings.

The overall $\mstar - \mbh$ relation in \astrid shows broad agreement compared to the scaling relation fit to observations of the local AGN population. 
Massive galaxies with $\mstar > 3 \times 10^{11} \msun$ typically host BHs with $\mbh \gtrsim 5 \times 10^{8} \msun$.
The host galaxies of the triple quasars at $z \geq 2.5$ have $\mbh - \mstar$ values comparable to (albeit slightly lower than) the overall distribution. 
All three systems are rich in gas and have high star formation rates $\sim 10^3 \msun$/yr at $z=2.7$. 
At $z=2.5$ when the first two galaxies (hosts of BH1 and BH2) begin their encounter and merge with each other, the central most massive galaxy experiences a starburst with star formation rate $> 5 \times 10^3 \msun$/yr which quickly builds up the stellar mass of the remnant. 
As illustrated by the inset panel, the BH1 host galaxy at $z=2.5$ exhibits a disturbed morphology during the merger process. 
At $z=2.3$, the merger of the second system is responsible for the assembly of the most massive galaxy in the simulation with $\mstar > 2 \times 10^{12} \msun$ and also leads to the most massive black hole with $\mbh \sim 10^{11} \msun$. 
The resultant $\mbh$ and $\mstar$ values for this system sit on the extrapolation of the scaling relation from \cite{Reines2016}.
The remnant galaxy at $z=2.0$ has a stellar mass of $\mstar = 2.3 \times 10^{12} \msun$, with half mass radius $r_{1/2} = 3$ pkpc, in agreement with the observations of $z>2$ massive galaxies finding that their morphologies are usually compact \citep[e.g.][]{Papovich2005ApJ...631..101P}.

In Figure~\ref{fig:Scaling-Relation} we show the total star formation rate as a function of stellar mass for the whole population of galaxies in the simulation. 
We follow the evolution of the star formation rate of the triple system and show in the inset images the star-forming gas surrounding BH1 before and after the merger.
Here we see how the starburst in the early stages of the galaxy merger produces a large star formation rate in the center of the galaxy.
The powerful AGN feedback is able to blow the star-forming gas out of the galaxy \citep[see also][on quasar outflow]{Ni2018MNRAS.481.4877N}, suppressing star formation in the central regions of the remnant host galaxy.
At this point, all of the remaining star formation (still of the order of $100 \msun$/yr) occurs in dense clouds on the outskirts of the galaxy (as shown by the image at $z\sim 2$). 
The star formation rate in the innermost 50 $\hkpc$ region from the central black hole is $\sim 10 \msun$/yr. 
The remnant of this triple quasar system is reminiscent of the observation of the large clouds of gas surrounding the quasar in the TON 618 system. 
ALMA has recently confirmed that the host galaxy of this system is surrounded by an enormous Lyman-$\alpha$ blob (LAB) with inferred molecular gas content sufficient to provide $\sim 50-100 \msun$/ yr at hundreds of kpc from the quasar itself \citep{Li2021ApJ...922L..29L}.
In the simulation, the star-forming gas pushed out by the quasar outflow, and the remnant $z \sim 2$ UMBH appears consistent with the observed environment for candidate UMBH in TON618 \citep{Li2021ApJ...922L..29L}.
This supports a scenario in which the formation of an UMBH involves a strong quasar phase (major triple merger) hosted by a strongly quenched host galaxy with associated molecular gas outflow and extended LAB system at $z\sim 2$.




\section{Discussion}
\label{section4:Discussion}


The most massive UMBH with $\mbh = 1.8 \times 10^{11} \msun$ at $z=2$ in \astrid is close to the theoretical estimate of the black hole mass upper limit, which is in the mass regime of $5 \times 10^{10} \sim 2.7 \times 10^{11} \msun$ \citep{King2016MNRAS.456L.109K}.
At $z=2$, there are two other UMBHs with $\mbh > 5 \times 10^{10} \msun$ in \astrid, with the second most massive black hole having mass $\mbh = 6.4 \times 10^{10} \msun$. 
We trace their evolution history and find that they have both experienced two massive black hole mergers involving $\mbh \sim 10^8 \msun$ black holes following the merger of their host galaxies.
Resembling the $10^{11} \msun$ UMBH, the other systems gained the majority of their black hole mass through active gas accretion induced by mergers of galaxies (with smaller stellar mass). 
Their progenitors are seen to be quasar triplets with lower luminosities of $L_{\rm bol} > 10^{45}$ ergs/s.

However, a caveat is that the UMBH mass seen in \astrid is not a prescription for a new upper limit for the black hole mass, as cosmological simulations cannot directly resolve the physical processes of black hole accretion below kpc scales. 
The growth or accretion state of black holes (and therefore the final black hole mass) might vary depending on specific prescriptions for the black hole sub-grid model that links the thermal state of gas on kpc scales to activity on unresolved scales. 
In this work, we instead show a scenario for how the most massive UMBH can be formed through succeeding mergers of SMBHs residing in massive, gas-rich galaxies that can be observed as quasar triplet systems.

Given the unprecedently large volume of \astrid, this may be the first time that such an extreme merger of three $M_* > 3 \times 10^{11} \msun$ galaxies at $z \sim 2$ has been directly modelled in a uniform volume cosmological hydrodynamic simulation.
We have established that a $10^{11} \msun$ UMBH can form in such an extreme event. 
The resultant $\mbh - \mstar$ relation lies on (or slightly above) the extrapolation of the $\mbh - \mstar$ fit based on local AGN observations. 
The UMBH resides at the center of a $M_{\rm h} = 10^{14} \msun$ halo at $z=2$. 
The number density of such massive halos is $\phi \sim 10^{-6}$ cMpc$^{-3}$, as rare as the massive clusters in the local universe with $M_{\rm h} \sim 2\times 10^{15} \msun$.
This implies that  $\mbh > 10^{11} \msun$ UMBHs may reside in the massive clusters in the local universe and have assembled their mass at cosmic noon through frequent mergers of massive galaxies.





\section{Conclusion}
\label{section5:Conclusion}

\astrid is the first cosmological hydrodynamic simulation with a volume large enough to cover a handful of the rare most massive $\mstar > 10^{12} \msun$ galaxies and $M_{\rm h} \sim 10^{14} \msun$ halos at the $z \sim 2$ epoch known as cosmic noon.
It models black hole growth and co-evolution with galaxies using a set of subgrid physical models that are in broad agreement with various observational constraints, such as galaxy populations, quasar luminosity functions, $\mbh - \mstar$ relations etc.
We find that ultramassive black holes with extreme masses of $\mbh > 5 \times 10^{10} \msun$ can be formed in the rare events that are multiple massive galaxy mergers happening around $z \sim 2$, the epoch when both star formation and AGN reach their peak activity.

We investigate the population of dual and triple quasars with luminosity threshold $L_{\rm bol} > 10^{45.5}$ ergs/s. For a given quasar at $z = 2 \sim 3$, the probability of finding another (dual) or two other (triple) quasars within a galactic separation of $5 \hkpc < r < 200 \hkpc $ (0.3\arcsec - 10\arcsec at $z \sim 2$) is on order of $f_{\rm QQ} \sim 1 \times 10^{-2}$ and $f_{\rm QQQ} \sim 2 \times 10^{-4}$.

We showed the formation of an UMBH with $\mbh \sim 10^{11} \msun$ at $z=2.3$, produced by the merger of a bright triplet quasar system with members having $L_{\rm bol} > 10^{46}$ ergs/s and residing in massive galaxies with $\mstar > 3 \times 10^{11} \msun$. 
The subsequent massive galaxy merger triggered active star formation with SFR $\sim 5 \times 10^{3} \msun$/yr and close to Eddington accretion of the central $10^{10} \msun$ black hole for about 140 Myrs, leading to an extreme quasar luminosity of $L_{\rm bol} > 10^{48}$ ergs/s (comparable to the most luminous quasar ever observed).
The merger formed a massive compact galaxy with $\mstar > 2 \times 10^{12} \msun$.
Powerful feedback from the UMBH quenched the star formation in the surroundings to $< 10 \msun$/yr in the innermost $50 \hkpc$ region.
For massive galaxies with $\mstar > 10^{12} \msun$, \astrid predicts $\mbh - \mstar$ on the extrapolation of the scaling relation fit to the observations of local AGN, showing that they can host UMBHs with $\mbh > 5 \times 10^{10} \msun$ through major galaxy mergers that induce the most active black hole growth.



\section*{Acknowledgements}
TDM and RACC acknowledge funding from the NSF AI Institute: Physics of the Future, NSF PHY-2020295, NASA ATP NNX17AK56G, and NASA ATP 80NSSC18K101.
\texttt{Astrid} was run on the Frontera facility at the Texas Advanced Computing Center.
We acknowledge Stephen Wilkins, Patrick Lachance, Yue Shen for useful discussions. 
TDM acknowledges additional support from  NSF ACI-1614853, NSF AST-1616168, NASA ATP 19-ATP19-0084, and NASA ATP 80NSSC20K0519, and RACC from NSF AST-1909193.
SB acknowledges funding supported by NASA-80NSSC21K1840.

\section*{Data Availability}
The code to reproduce the simulation is available at \url{https://github.com/MP-Gadget/MP-Gadget}, and continues to be developed.
Part of the \astrid snapshots are available at \url{https://astrid-portal.psc.edu/}.

\bibliography{bib}{}
\bibliographystyle{aasjournal}

\end{document}